%% file: MethodsComparisonSpikeSorting_arXiv.tex
\documentclass[10pt]{article}

\usepackage{amsmath}
\usepackage{graphicx}
\usepackage{fancyref}
\usepackage{hyperref}
\usepackage{booktabs}
\usepackage{authblk}
\usepackage[numbers,sort&compress]{natbib}
\usepackage[margin=2cm]{geometry}

\begin{document}

\title{Comparison of feature extraction and dimensionality reduction methods for single channel extracellular spike sorting}

\author[a]{Anupam Mitra \footnote{Email: anupam.mitra@gmail.com}}
\author[a]{Anagh Pathak \footnote{Email: pathak02@gmail.com}}
\author[a]{Kaushik Majumdar \footnote{Email: kmajumdar@isibang.ac.in}}
\affil[a]{Systems Science and Informatics Unit, Indian Statistical Institute, 8th Mile, Mysore Road, Bangalore 540059, India}

\maketitle

\begin{abstract}
\input{Abstract}

\end{abstract}
\input{Body}

\section*{Acknowledgements}
The work has been supported by the Indian Statistical Institute through an internal grant no. SSIU-03O-2014-16 and by the Department of Biotechnology, Government of India through grant no. BT/PE7666/MED/30/736/2013. We thank the authors of \cite{quiroga2004unsupervised} and \cite{pedreira2012many} for making the data used in their work available, which we used in this work. We also thank Prof Kalyan Chowdhury and Aditya Ramesh for useful discussions.

\bibliographystyle{unsrt}
\bibliography{references}{}

\end{document}

%% file: Abstract.tex
Spikes in the membrane electrical potentials of neurons play a major role in the functioning of nervous systems of animals. Obtaining the spikes from different neurons has been a challenging problem for decades. Several schemes have been proposed for spike sorting to isolate the spikes of individual neurons from electrical recordings in extracellular media. However, there is much scope for improvement in the accuracies obtained using the prevailing methods of spike sorting. To determine more effective spike sorting strategies using well known methods, we compared different types of signal features and techniques for dimensionality reduction in feature space. We tried to determine an optimum or near optimum feature extraction and dimensionality reduction methods and an optimum or near optimum number of features for spike sorting. We assessed relative performance of well known methods on simulated recordings specially designed for development and benchmarking of spike sorting schemes, with varying number of spike classes and the well established method of $k$-means clustering of selected features. We found that almost all well known methods performed quite well. Nevertheless, from spike waveforms of 64 samples, sampled at 24 kHz, using principal component analysis (PCA) to select around 46 to 55 features led to the better spike sorting performance than most other methods (Wilcoxon signed rank sum test, $p < 0.001$).

%% file: Body.tex
\section{Introduction}
Action potentials are large fluctuations in the membrane potential which travel over long distances at relatively high speeds \cite{dayan2001theoretical, koch1998biophysics, quiroga2007spike, gibson2012spike}. These action potentials, also called spikes, are a major mode of communication between neurons in an animal's nervous system \cite{dayan2001theoretical, koch1998biophysics, quiroga2007spike, gibson2012spike, einevoll2012towards}. Therefore, studies of the brain function at cellular level requires detection of spikes from as many neurons as possible \cite{rey2015past, einevoll2012towards}. Isolating the activity of individual neurons in the signal greatly improves our understanding of the role of each individual neuron in the function of the brain and rest of the nervous system \cite{buzsaki2004large, quiroga2005invariant, einevoll2012towards, rey2015past}. Moreover, it is also important for brain computer interface (BCI) \cite{obeid2004evaluation, serruya2002brain} and in clinical treatment and research studies of epilepsy patients \cite{truccolo2011single}.

Several techniques have been used for measuring the activity of individual neurons in the brain \cite{gerstein1964simultaneous, buzsaki2004large, baker1999multiple, stevenson2011advances, pouzat2014spysort, einevoll2012towards}. Small electrodes have been used to record electrical signals from neurons \cite{gerstein1964simultaneous, buzsaki2004large, baker1999multiple, stevenson2011advances}. Cellular electrical signals can be recorded from within a cell (intracellular) or from outside a cell (extracellular) \cite{wehr1999simultaneous, harris2000accuracy, buzsaki2004large}. Small diameter electrodes are placed in the intracellular medium to record activity of the cell or the extracellular medium to record voltage signals from several nearby cells. While an extracellular recording provides information about the electrical activity of several cells, in practice isolating the activity of individual cell is quite challenging \cite{lewicki1998review, quiroga2004unsupervised, quiroga2007spike, gibson2012spike, pedreira2012many, rey2015past}. An electrode may have a single wire or multiple wires - two wire stereotrodes \cite{mcnaughton1983stereotrode, pouzat2014spysort}, four wire tetrodes \cite{gray1995tetrodes, harris2000accuracy, pouzat2014spysort} or more \cite{pouzat2014spysort, einevoll2012towards}. The challenges in isolating spikes from single wire electrodes and multiple wire electrodes are somewhat different \cite{rossant2015spike}. In this paper, we focus exclusively on finding the best way to isolate spikes from single wire electrodes, which are often used in cellular neural recordings, especially with human subjects \cite{quiroga2005invariant}.

Spike sorting is the process of isolating spikes from different neurons from voltage signals recorded from extracellular media \cite{lewicki1998review, brown2004multiple, quiroga2007spike, gibson2012spike, rey2015past, pouzat2014spysort, einevoll2012towards}. It is particularly challenging in an population of neurons with high packing density \cite{buzsaki2004large}. Many schemes been developed for spike sorting, using a pipeline of methods derived from signal processing, machine learning and statistics \cite{lewicki1998review, quiroga2004unsupervised, quiroga2007spike, gibson2008comparison, gibson2010technology, gibson2012spike, pedreira2012many, paraskevopoulou2013feature, pouzat2014spysort, einevoll2012towards, rey2015past}. However, there is much scope for improvement for spike sorting accuracy \cite{pedreira2012many, gibson2012spike}. Furthermore, it is not known what combination of methods leads to the best performance. A comparison of popular spike sorting software like WaveClus \cite{quiroga2004unsupervised}, OSort \cite{rutishauser2006online} and Klustakwick \cite{harris2000accuracy} has been done in \cite{wild2012performance}, treating each program as a black box with tunable parameters. In this work, we compared the techniques used for different stages of spike sorting directly by benchmarking them on well known simulated data designed to determine the performance of spike sorting.

The extracellular signal contains a low frequency component, the Local Field Potential (LFP), which is primarily due to synaptic activity and a high frequency component due to spikes \cite{buzsaki2012origin, quiroga2004unsupervised, quiroga2007spike}. Typically, the extracellular signal is low pass filtered within around 300 Hz to obtain the LFP, and high band pass filtered between approximately 300 Hz to 3 kHz to obtain the signals due to spiking activity of neurons \cite{buzsaki2012origin, quiroga2004unsupervised, quiroga2007spike}. The high pass filtered signal is used for spike sorting. The basic spike sorting process involves detection of spikes, spike alignment, feature extraction, dimensionality reduction and clustering \cite{quiroga2004unsupervised, quiroga2007spike, gibson2008comparison, gibson2010technology, gibson2012spike, rey2015past}. The final clustering step may be replaced by template matching to existing templates if templates are available \cite{quiroga2004unsupervised, quiroga2007spike, rey2015past}. Often in practice the spike sorting system switches to template matching after performing clustering on a few thousand spikes \cite{quiroga2004unsupervised, quiroga2007spike, rey2015past}. This is usually followed by post processing by looking at the interspike intervals to determine whether each cluster corresponds to a single unit or a multi unit \cite{quiroga2004unsupervised, quiroga2007spike, rey2015past}. Towards an optimum spike sorting scheme, we tried to determine which methods work the best for spike feature extraction and dimensionality reduction by comparing the performance of different schemes used for these steps, keeping the other steps fixed. In other words among well known methods, we wanted to answer the following questions.

\begin{enumerate}
\item Which feature extraction method leads to the best spike sorting?
\item Which dimensionality reduction method leads to the best spike sorting?
\item What is the optimum number of features to use for spike sorting?
\end{enumerate}

The rest of the paper is as follows. In Section \ref{sec:Methods}, we describe the feature extraction and the dimensionality reduction techniques we used. In Section \ref{sec:Data}, we describe the data used for our analysis. In Section \ref{sec:Results}, we elaborate on our results and try to answer the above questions. Finally in Section \ref{sec:Discussion}, we conclude with a discussion of our work and suggest possible future directions.

\section{Methods}
\label{sec:Methods}
Conventional spike sorting involves a pipeline of several steps, including spike detection, feature extraction, dimensionality reduction and clustering, that is the following steps \cite{lewicki1998review, quiroga2004unsupervised, quiroga2007spike, gibson2008comparison, gibson2010technology, gibson2012spike, pedreira2012many, paraskevopoulou2013feature, rey2015past}.

\begin{enumerate}
\item Band pass filtering the signal \cite{quiroga2004unsupervised, quiroga2007spike, gibson2008comparison, gibson2010technology, gibson2012spike, pedreira2012many, pouzat2014spysort, rey2015past}
\item Detection of spikes \cite{lewicki1998review, quiroga2004unsupervised, quiroga2007spike, gibson2008comparison, gibson2010technology, gibson2012spike, pedreira2012many, paraskevopoulou2013feature, pouzat2014spysort, rey2015past}
\item Extraction and alignment of spike waveforms \cite{quiroga2004unsupervised, gibson2008comparison, gibson2010technology, gibson2012spike, pedreira2012many, paraskevopoulou2013feature, pouzat2014spysort, rey2015past}
\item Extraction of spike features \cite{quiroga2004unsupervised, gibson2008comparison, gibson2010technology, gibson2012spike, pedreira2012many, paraskevopoulou2013feature, rey2015past, bestel2012novel, pouzat2014spysort}
\item Selection of spike features \cite{quiroga2004unsupervised, gibson2008comparison, gibson2010technology, gibson2012spike, pedreira2012many, paraskevopoulou2013feature, pouzat2014spysort, rey2015past}
\item Clustering using spike features \cite{quiroga2004unsupervised, gibson2008comparison, gibson2010technology, bestel2012novel, gibson2012spike, pedreira2012many, paraskevopoulou2013feature, pouzat2014spysort, rey2015past}
\item Post processing using interspike interval distributions to determine whether a cluster obtained in the previous step is a single unit cluster or a multi unit cluster \cite{quiroga2004unsupervised, gibson2012spike, pedreira2012many, rey2015past}
\end{enumerate}

In this paper we focus on extraction of spike features and selection of an optimum or near optimum set of spike features. We assumed that the detection is perfect and extracted spike waveforms based on the known spike times. For our analysis, we extracted a 2.5 millisecond waveform, consisting of 64 samples at a sampling rate of $f_{\text{samp}}$ = 24 kHz from each spike, as done in \cite{quiroga2004unsupervised}. For each waveform, we extracted different types of features. Thereafter, we used different dimensionality reduction techniques to select between 1 to 64 features. Using the selected features, we performed clustering to estimate the spikes which appeared to belong to the same neuron.

Commonly used methods for feature extraction are numerical differentiation \cite{paraskevopoulou2013feature, gibson2008comparison, gibson2010technology, gibson2012spike} wavelet decomposition \cite{quiroga2004unsupervised, geng2012unsupervised} and principal component analysis (PCA) \cite{lewicki1998review, quiroga2004unsupervised, bestel2012novel}. Using features based on the raw waveform, numerical differentiation and Haar wavelet decomposition, we obtain about as many features as the number of samples in the spike waveforms. It is not clear which subset(s) of these features would lead to the best spike sorting. Several techniques like maximum variance, maximum difference and Lilliefors test have been used as heuristics to select a small subset of features \cite{gibson2008comparison, quiroga2004unsupervised, quiroga2007spike}. When features are obtained using PCA, they can be ranked based on their corresponding eigenvalues and the features with the highest eigenvalues can be chosen \cite{quiroga2004unsupervised, quiroga2007spike}. Therefore, we used PCA as a dimensionality reduction step, after feature extraction.

We wanted to find an optimum or near optimum feature extraction method, dimensionality reduction method and number of features for spike sorting. We considered four types of feature extraction methods based on the raw spike waveforms, Haar wavelet decomposition \cite{quiroga2004unsupervised}, first and second order differences \cite{paraskevopoulou2013feature} and first order differences with lag \cite{gibson2008comparison}. We also used three methods of dimensionality reduction using maximum variance, Lilliefors test and PCA. Note that we consider PCA to be a dimensionality reduction process, rather than a feature extraction process, as is commonly done in the spike sorting literature. Using these we obtained twelve methods for feature extraction and selection, which are summarized in Table \ref{tab:MethodsDescription} and discussed in the following paragraphs.

\subsection{Features}
For dimensionality reduction, we used four types of features based on the raw spike waveform, Haar wavelet decomposition \cite{quiroga2004unsupervised}, first and second order differences \cite{paraskevopoulou2013feature} and first order differences with lag \cite{gibson2008comparison}. These gave 64 or more features, which were used as an input to the dimensionality reduction step.

\subsubsection{Raw waveform based features}
Features based on the raw signal were simply samples from the spike waveform. Using a 2.5 millisecond waveform sampled at 24 kHz, we obtained 64 samples, all of which we took as features. Therefore the feature vector was 64 dimensional.

\subsubsection{First and second difference based features}
Using a finite difference approximation to the first and second time derivatives of the spike waveform has been shown to be an effective way of extracting features for spike sorting \cite{paraskevopoulou2013feature}. Using forward differences between adjacent samples in the signal, we obtain the simplest finite difference approximation to the first time derivative of the signal, $s'[n]$.
\begin{equation}
\label{eq:FirstDifference}
s'[n] = s[n+1] - s[n]
\end{equation}

Using forward differences between adjacent samples of the forward first difference approximation $s'[n]$, we obtain the simplest finite difference approximation to the second time derivative of the signal, $s''[n]$.
\begin{equation}
\label{eq:SecondDifference}
s''[n] = s'[n+1] - s'[n]
\end{equation}

Using Eq (\ref{eq:FirstDifference}) and (\ref{eq:SecondDifference}) on a 64 sample window, we obtain signals with 63 and 62 samples respectively. We concatenated them to obtain 125 features.

\subsubsection{First difference with lag based features}
Forward differences similar to Eq (\ref{eq:FirstDifference}), but with gap of more than one sample in the difference have also been shown to be quite effective for spike sorting \cite{gibson2008comparison, gibson2010technology, gibson2012spike}. We use a gap of $k$ points to compute the first difference with a lag of $k$ points. 
\begin{equation}
s'[n] = s[n+k] - s[n]
\end{equation}

Following \cite{gibson2008comparison, gibson2010technology, gibson2012spike}, where these features have been called ``discrete derivatives'', we used $k \in \{1, 3, 7\}$. This gave us 63, 61 and 57 features respectively from a 64 sample window. We concatenated these to obtain 181 features.

\subsubsection{Haar wavelet decomposition based features}
Wavelet decomposition is quite popular for spike sorting \cite{quiroga2004unsupervised, geng2012unsupervised}. Wavelet decomposition involves decomposing the signal in terms of contracted/dilated and translated version of a ``mother wavelet'' function. This decomposition is obtained by convolving these wavelet functions to obtain coefficients describing the signal in terms of wavelet basis functions. Each coefficient describes how well the signal can be locally approximated by the corresponding wavelet basis function. Since the wavelet basis functions vary in duration and position, they provide information about the signal at different time scales at different times \cite{quiroga2004unsupervised}. Following \cite{quiroga2004unsupervised}, we used Haar wavelets as they have a finite support and the Haar wavelet basis functions are orthogonal.

\subsection{Dimensionality Reduction}
For selection of features, we use three dimensionality reduction techniques: maximum variance, principal component analysis and Lilliefors test for normality. We ranked features using these criteria and selected the best $m$ features, varying $m$ from 1 to 64. Some types of feature extraction methods give more than 64 features, but we did not select more than 64 features as the raw spike waveforms have 64 samples in our analysis.

\subsubsection{Maximum variance}
Features which have higher variance can help distinguish one class of spikes from another. Based on this hypothesis, we ranked each feature using its variance, from the highest variance to the lowest variance. We selected the best $m$ features from this ranking in the dimensionality reduction step.

\subsubsection{Lillifors test for normality}
Another approach to selecting the best set of features is to use the Lilliefors test for normality to find the coefficients whose distribution is the most multimodal \cite{quiroga2004unsupervised, quiroga2007spike, rey2015past}. The underlying assumption is that the more multimodal features will lead to better discrimination of spikes of different classes \cite{quiroga2004unsupervised, quiroga2007spike, gibson2012spike}.

The Lilliefors test is a modification of the Kolmogorov Smirnov test, which compares the distribution of a random variable to a normal distribution. The larger the Lillifors test statistic, the more different is the distribution of the random variable (which in our case is the feature being considered) from the normal distribution. The deviance from a normal distribution is assumed to be correlated with multimodality \cite{quiroga2004unsupervised, quiroga2007spike, rey2015past}.

We ranked feature using the Lilliefors test statistic, from highest to lowest and selected the best $m$ features from this ranking as the dimensionality reduction step.

\subsubsection{Principal component analysis}
Principal component analysis (PCA) is the process of obtaining a set of orthogonal feature vectors, called principal components, which capture the directions of variation in the data, in decreasing order \cite{shlens2014tutorial}. Usually, taking the vectors corresponding to the first few principal components suffices in capturing most of the variance in the data. The first few principal components have widely been used as a dimensionality reduction step for spike sorting \cite{lewicki1998review, quiroga2004unsupervised, adamos2008performance, gibson2008comparison, pouzat2014spysort}.

In our work, we consider PCA to be a dimensionality reduction step to select a smaller number of features from the features extracted using methods described earlier. Therefore in our analysis, method 1 (see Table \ref{tab:MethodsDescription}) which involves feature extraction from the raw waveform and dimensionality reduction using PCA is equivalent to the typical practice of using PCA as a feature extraction step. For dimensionality reduction from each class of features, we used PCA to obtain the best $m$ principal components as features. It should be noted that unlike in other dimensionality reduction methods (like maximum variance and Lilliefors test), PCA does not select a subset of the original features, but linear combinations of the original features.

\subsection{Clustering}
Using four types of features and three criteria for dimensionality reduction, we got twelve methods, which are summarized in Table \ref{tab:MethodsDescription}. After obtaining the features, we established clusters of spikes based on similarity in their selected features. Among clustering algorithms, $k$-means has been found useful for benchmarking features \cite{gibson2012spike, hulata2002method, paraskevopoulou2013feature}. A drawback of using $k$-means clustering is that it requires the number of clusters, $k$, to be known, a priori. While this may not be feasible in an actual experimental scenario, where the number of different neurons contributing to the extracellular signal is unknown, this approach does help in benchmarking the performance of features \cite{gibson2012spike, paraskevopoulou2013feature}. Another drawback is $k$-means is based on an Euclidean distance based similarity measure, which may not hold for spike features, especially in scenarios with electrode drift and non Gaussian distributions of spike features \cite{quiroga2004unsupervised, quiroga2007spike, gibson2012spike}. Nevertheless, $k$-means has been found to give the maximum clustering accuracy among most commonly used clustering algorithms used for spike sorting \cite{karkare201375}.

\begin{table}
\centering
\begin{tabular}{rll}
 & \textbf{Feature} & \textbf{Dimensionality} \\
 & \textbf{Extraction} & \textbf{Reduction} \\
\toprule
1. & Raw & PCA \\
2. & Raw & Var \\
3. & Raw & LT \\
4. & HW & PCA \\
5. & HW & Var \\
6. & HW & LT \\
7. & FSD & PCA \\
8. & FSD & Var \\
9. & FSD & LT \\
10. & FDL & PCA \\
11. & FDL & Var \\
12. & FDL & LT \\
\end{tabular}
\caption{Methods for feature extraction and dimensionality reduction. The abbreviations used in the table are as follows: Var = Maximum Variance, LT = Lilliefors Test, PCA = Principal Component Analysis, HW = Haar wavelet decomposition, FSD = First and second difference and FDL = First difference with lag.}
\label{tab:MethodsDescription}
\end{table}

\subsection{Implementation}
We implemented all methods in Python using the NumPy \cite{van2011numpy}, Scipy \cite{jones2014scipy} and Scikit-learn libraries \cite{scikit-learn}. $k$-means clustering was performed using the randomized $k$-means algorithm implemented in the Scikit-learn library \cite{scikit-learn}. Haar Wavelet decomposition was done using the PyWavelets library \cite{wasilewski2010pywavelets}.

\section{Data}
\label{sec:Data}
Using simulated datasets, where the ground truth for each spike is known enables us to assess the performance of different spike sorting schemes. Therefore, we used the data used in \cite{pedreira2012many} for performance evaluation of different spike sorting schemes. These data and similarly simulated data were produced for the purpose of determining the performance of different spike sorting schemes \cite{quiroga2004unsupervised, martinez2009realistic} and have been used extensively for this purpose \cite{quiroga2004unsupervised, gibson2008comparison, martinez2009realistic, gibson2010technology, gibson2012spike, pedreira2012many, wild2012performance, paraskevopoulou2013feature}

Synthetic extracellular recordings modelling the contribution of the background noise, multi-unit and single-unit activity were created using a database with 594 different averaged spike shapes, taken from real recordings from neocortex and basal ganglia \cite{quiroga2004unsupervised, martinez2009realistic}. The data were first simulated at a sampling frequency of 96 kHz, then they were down sampled to 24 kHz \cite{quiroga2004unsupervised, martinez2009realistic}.

The data were produced with different number of spike classes in order to estimate the number of neurons whose spikes can be isolated from an extracellular recording \cite{pedreira2012many}. Simulations as described earlier were used to create 10 minute long recordings, containing between 2 and 20 different classes of spikes. For each number of spike classes there were 5 recordings, giving a total of 95 recordings.

\section{Results}
\label{sec:Results}

\begin{table}
\centering
\begin{tabular}{lr}
\toprule
\textbf{Feature} & \textbf{Fraction of} \\
\textbf{Extraction} & \textbf{best performances} \\
\toprule
Haar wavelet decomposition & $24/95 \approx 0.253$ \\
First and second differences & $18/95 \approx 0.189$ \\
First difference with lag & $26/95 \approx 0.274$ \\
Raw waveform & $27/95 \approx 0.284$ \\
\bottomrule
\end{tabular}
\caption{The number of recordings in which each class of features performed the best in the data of \cite{pedreira2012many}. No specific feature class stand out as significantly better performing than others.}
\label{tab:FeatureClassBestPerform}
\end{table}

\begin{table}
\centering
\begin{tabular}{lr}
\toprule
\textbf{Dimensionality} & \textbf{Fraction of} \\
\textbf{Reduction} & \textbf{best performances} \\
\toprule
Principal component analysis & $79/95 \approx 0.831 $ \\
Maximum variance & $15/95 \approx 0.158$ \\
Lilliefors test & $01/95 \approx 0.011$ \\
\bottomrule
\end{tabular}
\caption{The number of recordings in which each dimensionality reduction criterion performed the best in the data of \cite{pedreira2012many}. PCA performed best on most of the recordings.}
\label{tab:FeatureSelectCriterionBestPerform}
\end{table}

\begin{figure*}
\centering
\includegraphics[scale=1.0]{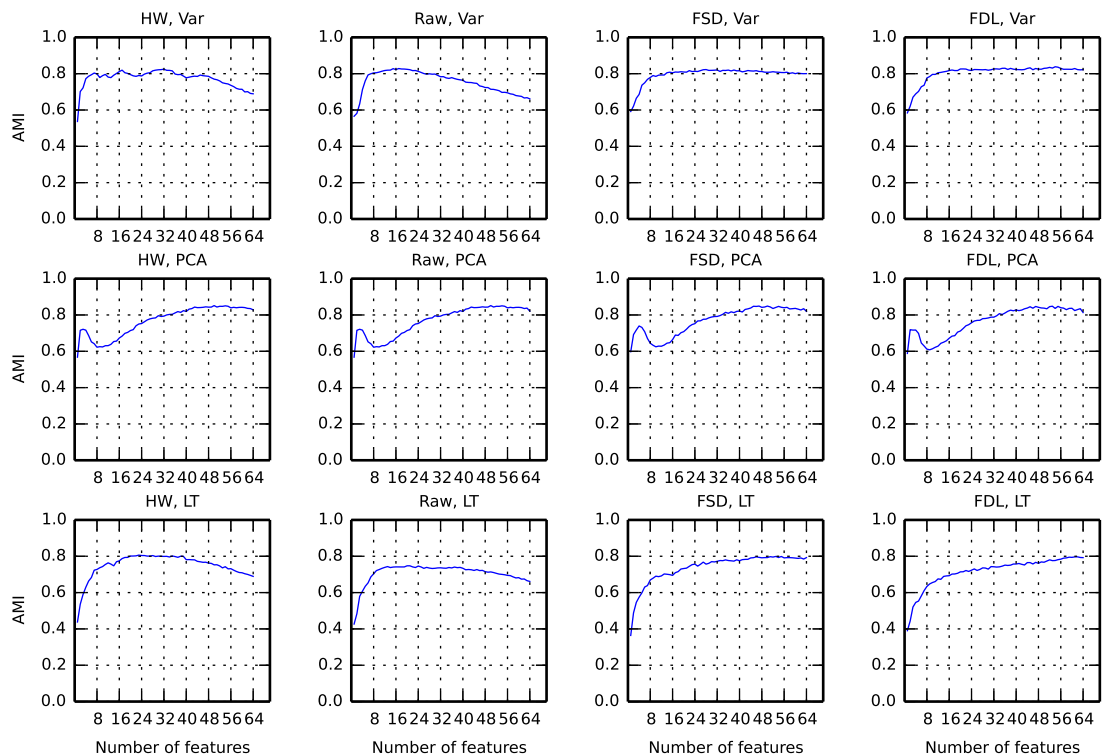}
\caption{Median AMI scores as the number of features is increased for different class of features on the data of \cite{pedreira2012many}. Qualitatively, we see that using PCA for the dimensionality reduction step gives the best AMI scores for all types of features, when the number of features is around 48. However using less than 24 features, PCA seems to perform much worse than other methods. This is consistent with the belief that the first few principal components do not correspond to features that discriminate well between spikes \cite{quiroga2004unsupervised, quiroga2007spike, rey2015past}, considering that typically about 10 features are used for spike sorting \cite{quiroga2004unsupervised, quiroga2007spike, gibson2010technology}. For a detailed discussion see Subsection \ref{subsec:SpikeSortingResults}. The abbreviations used in the figure are as follows: Var = Maximum Variance, LT = Lilliefors Test, PCA = Principal Component Analysis, HW = Haar wavelet decomposition, FSD = First and second difference and FDL = First difference with lag.}
\label{fig:Comparison}
\end{figure*}

\begin{figure*}
\centering
\includegraphics[scale=0.9]{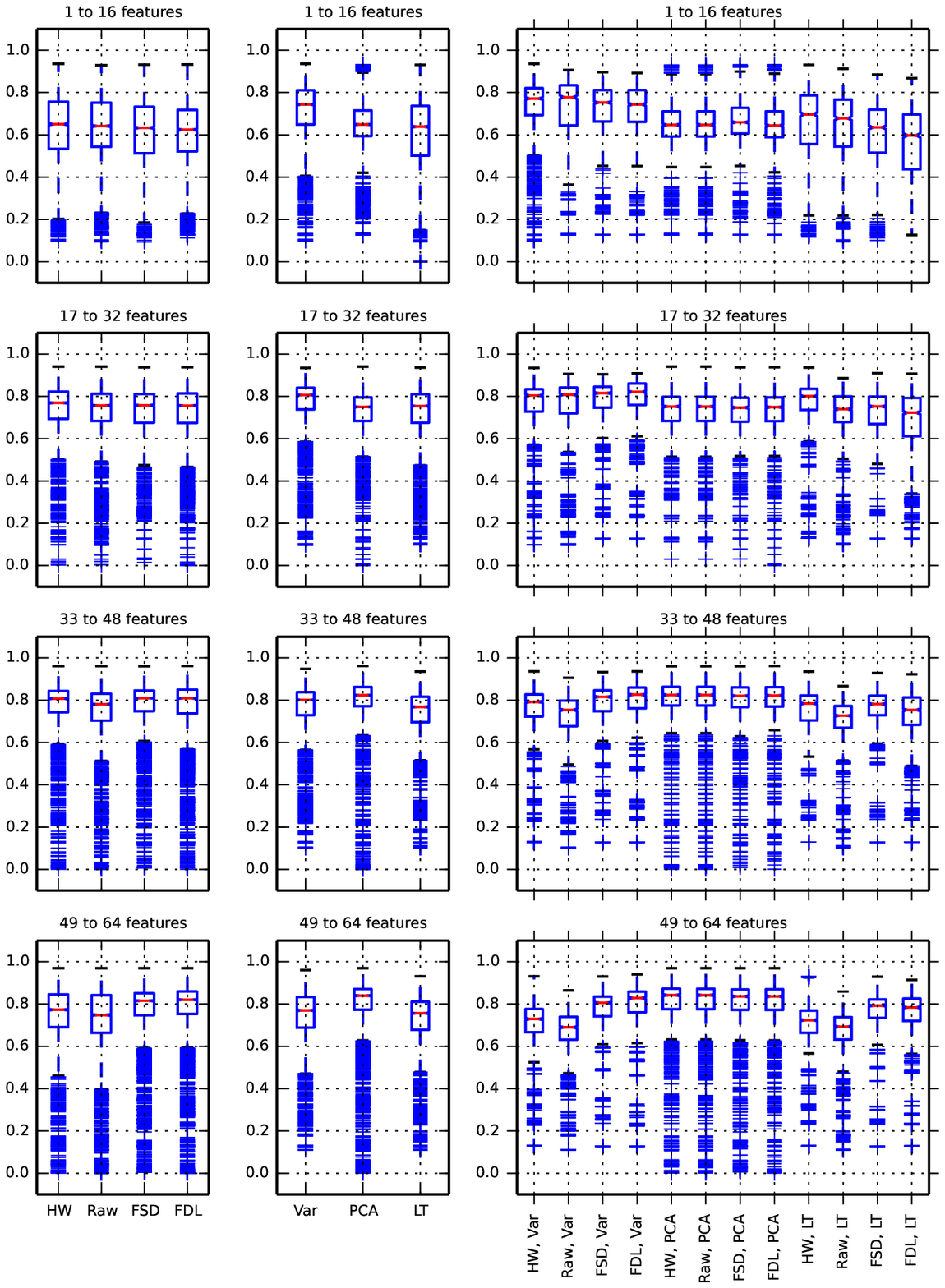}
\caption{Box plots of AMI scores for different number of features on the data of \cite{pedreira2012many}. The first column shows the AMI scores of different feature extraction methods for different number of features. The second column shows the AMI scores of different dimensionality reduction methods for different number of features. The third column shows the AMI scores for different pairs of feature extraction and dimensionality reduction methods. For a discussion see Subsection \ref{subsec:SpikeSortingResults}.
The abbreviations used in the figure are as follows: Var = Maximum Variance, LT = Lilliefors Test, PCA = Principal Component Analysis, HW = Haar wavelet decomposition, FSD = First and second difference and FDL = First difference with lag.}
\label{fig:Boxplots}
\end{figure*}

An effective method of determine performance of spike sorting is to use simulated data where the ground truth is known  \cite{quiroga2004unsupervised, martinez2009realistic, gibson2008comparison, gibson2010technology, einevoll2012towards}. Using simulated recordings enables assessing performance by comparison with ground truth \cite{quiroga2004unsupervised, martinez2009realistic, gibson2008comparison, gibson2010technology, einevoll2012towards}. We used adjusted mutual information to compare the results of spike sorting with the known ground truth, following previous work \cite{gasthaus2008spike, kretzberg2009comparison, wild2012performance}.

\subsection{Adjusted Mutual Information}
Several measures have been introduced for comparison of two clustering schemes on the same data \cite{rand1971objective, vinh2009information, vinh2010information}. Some of these measures have been used to assess the performance of spike sorting schemes \cite{gasthaus2008spike, kretzberg2009comparison, wild2012performance}. It was shown that adjusted mutual information (AMI) is the most appropriate among these measures \cite{vinh2009information, vinh2010information, wild2012performance}.

Mutual information between two random variables quantifies the amount of information obtained about one random variable through the other random variable \cite{vinh2009information, vinh2010information, cover2012elements}. It is symmetric in the random variables. For a clustering scheme the cluster membership of each spike can be considered a random variable. We can compare the similarity of two clustering schemes by measuring the mutual information between them \cite{rand1971objective, vinh2009information, vinh2010information}. The adjusted mutual information is defined by adjusting for chance similarity by subtracting the expected value of mutual information and normalizing \cite{vinh2009information, vinh2010information}. AMI = 1 corresponds to perfect match between the clustering schemes being compared and AMI = 0 corresponds to a chance match.

\subsection{Spike Sorting Results}
\label{subsec:SpikeSortingResults}
The data used in \cite{pedreira2012many} consisted of recordings where the number of spike classes were different, varying between 2 and 20 (both inclusive). Using the results of different recordings in the data, we try to find an optimum or near optimum combination of feature extraction method, dimensionality reduction method and number of features for spike sorting.

For each recording, we used the AMI score to assess the similarities of different spike sorting schemes to the ground truth. Perfect spike sorting would give an AMI score of 1. We used median AMI scores (following \cite{wild2012performance}) as most of the distributions were skewed towards larger AMI scores. The evolution of median AMI scores as the number of features used for clustering is increased from 1 to 64, are shown in Fig. \ref{fig:Comparison}. Here we see that using about 48 to 56 features selected using PCA gives the best performance, for all types of features. From Fig. \ref{fig:Comparison}, we also see that for all types of features, when using PCA for dimensionality reduction, the AMI scores are quite close. However, while using Lilliefor's test or maximum variance criterion, the AMI scores vary significantly for different types of features as the number of features is increased.

For illustration, we divided the number of features into four groups: 1 to 16 features, 17 to 32 features, 33 to 48 features and 49 to 64 features. In each group, we compared the performance of each of the methods across the data. The results are summarized as box plots in Fig. \ref{fig:Boxplots}.

\begin{enumerate}
\item Using between 1 to 16 features, most of the feature extraction and dimensionality reduction methods perform well, (median AMI between 0.6 and 0.8). However, when PCA is used as a dimensionality reduction method, the performance is much worse (median AMI less than 0.7). The relatively poor performance of PCA in this regime is in accordance with the idea that the first few principal components do not correspond to features that discriminate well between spikes \cite{quiroga2004unsupervised, quiroga2007spike, rey2015past} which applies to the common practice of using about 10 features \cite{quiroga2004unsupervised, quiroga2007spike, gibson2010technology}.
\item Using between 17 to 32 features, the performance of most of the feature extraction and dimensionality reduction methods is even better (median AMI around 0.8). Again PCA as the dimensionality reduction method in this regime leads to relatively worse performance (median AMI between 0.6 and 0.8).
\item Using between 33 to 48 features, the performance of all feature extraction and dimensionality reduction methods is comparable and good (median AMI around 0.8). Using PCA as the dimensionality reduction method in this regime leads to performance comparable to other methods.
\item Using between 49 and 64 features, most the feature extraction and dimensionality reduction methods perform well (median AMI between 0.6 and 0.8). However, using PCA in this regime leads to better performance than other methods (median AMI more than 0.8).
\item The AMI scores have lesser variance when using more features than using less features.
\end{enumerate}

We compared the AMI scores obtained using all twelve methods (Table \ref{tab:MethodsDescription}) with upto 64 features, giving 768 different combinations, pairwise using the Wilcoxon signed rank sum test. We found that any feature extraction method followed by PCA to obtain between 42 to 55 features led to better AMI ($p < 0.001$) than more than 512/768 of the other methods. Using these combinations, we get a median AMI of around 0.83. If more than 56 features are used, the performance of any feature extraction method followed by PCA reduces compared to the previous case, but is still very good (median AMI around 0.8).

Dimensionality reduction using PCA gives the best performance in most of the recordings in the data (Table \ref{tab:FeatureSelectCriterionBestPerform}). Moreover, all feature extraction methods leads to a comparable number of best performances (Table \ref{tab:FeatureClassBestPerform}).

Interestingly, using too many features reduces the AMI scores. This is especially true when PCA, raw waveform and Haar wavelet decomposition are used (Fig. \ref{fig:Comparison}). The extra features (like the least dominant principal components or least dominant wavelet components) maybe those corresponding to noise and adversely affect the spike sorting process. Qualitatively similar conclusions in multi electrode spike sorting was the motivation behind using masking of some features before spike sorting \cite{kadir2014high}. 

If the number of computations is major concern, such as in an on chip real time spike sorting scenario, PCA may not be the most practical. In such a circumstance, we found using maximum variance criterion for dimensionality reduction to be quite effective. A similar conclusion was reached in \cite{gibson2008comparison, gibson2010technology, gibson2012spike}.

\section{Conclusion}
\label{sec:Discussion}
The problem of extracting spikes from different neurons from extracellular recordings remains challenging and several elaborate schemes using a chain of methods adapted from different fields have been used to tackle it. In this work, we looked at two of the steps in this chain of steps, specifically spike feature extraction and dimensionality reduction. We evaluated combinations of four feature extraction methods and three dimensionality reduction methods for single electrode spike sorting (Table \ref{tab:MethodsDescription}). Using adjusted mutual information as a performance metric, we found that dimensionality reduction using PCA gave the best results with all feature extraction methods, using 46 to 55 features from each 64 sample spike waveform. Besides PCA, almost all other methods also led to very good spike sorting performance. Using PCA with a few features led to relatively worse performance. Since the performance of spike sorting using PCA as a dimensionality reduction technique did not vary with the choice of class of features, we found using PCA on raw features to select about 46 to 55 features to be the most effective spike sorting scheme. In this scheme, the feature extraction step is merely extraction of the spike waveform. This approach, albeit with lesser features, was among the first approaches to spike sorting and is still considered a trusted workhorse for analyses where the number of computations required is not a major concern \cite{lewicki1998review, quiroga2004unsupervised, bestel2012novel, kadir2014high, pouzat2014spysort}. We also confirmed that PCA is less effective when about 10 features are used \cite{quiroga2004unsupervised, quiroga2007spike, rey2015past} (Fig. \ref{fig:Comparison} and Fig. \ref{fig:Boxplots}).

We used $k$ means for the clustering step of spike sorting to provide a comparison of different methods used before it. While $k$ means has been found to give the most accurate results, it has limitations for spike sorting - it assumes a spherical distribution of features with Cartesian geometries and it requires knowledge of the number of spike classes beforehand \cite{quiroga2004unsupervised, quiroga2007spike, gibson2012spike}. Our analysis can be extended to compare the performance of the different clustering methods, especially those that address some of these issues like OSort \cite{rutishauser2006online} and super paramagnetic clustering \cite{quiroga2004unsupervised} with different types of features and dimensionality reduction methods. The data we used did not address the challenges of electrode drift and overlapping spike waveforms \cite{pedreira2012many}. We hope that our findings can help towards effective solutions to these challenges. We did not explore the computational complexity of different methods as we want to find the best strategy, with reasonably tractable computational cost. We hope our observations can provide a direction to developing spike sorting strategies that maintain and improve the accuracies we observed and are relatively less computationally expensive.